# Scaling and Better Approximating Quantum Fourier Transform by Higher Radices

Zeljko Zilic and Katarzyna Radecka, *Member, IEEE*

*Abstract*— Quantum Fourier Transform (QFT) plays a principal role in the development of efficient quantum algorithms. Since the number of quantum bits that can currently built is limited, while many quantum technologies are inherently three- (or more) valued, we consider extending the reach of the realistic quantum systems by building a QFT over ternary quantum digits. Compared to traditional binary QFT, the $q$-valued transform improves approximation properties and increases the state space by a factor of $(q/2)^n$. Further, we use non-binary QFT derivation to generalize and improve the approximation bounds for QFT.

## I. INTRODUCTION

Q UANTUM computing [15] was so far demonstrated as being capable of speeding up algorithms for problems such as factoring integers or database searching, as well as being useful in cryptographic applications. Several working quantum-computing systems have already been demonstrated in practice [8], [17]. With such a prospect, it is natural to seek new algorithmic solutions that exploit quantum mechanical principles.

Some of the major improvements of quantum computing over the classical algorithms are due to *Quantum Fourier Transform* (QFT) [12]. While classical Fast Fourier Transform (FFT) [4] decreases running time from $O(2^{2n})$ to $O(n2^n)$, QFT offers even more dramatic improvements. Although more restrictive in scope than classical FFT, it can be performed in time $O(n^2)$ or even $O(n)$ in some cases [12]. Direct benefits of QFT are obtained in the celebrated Shor's number factoring algorithm [16]. It was further observed that the derivation of general quantum operations in non-binary quantum logic relies on QFT [9].

Currently, the main obstacle in quantum computing is posed by inability to realize practical quantum systems operating on a large number of quantum bits. Using multiple-valued instead of binary logic has a significant impact on the representable state space and the computational power of quantum machines. It is also worth noting that some known quantum circuit technologies, such as ion traps or quantum dots, are actually three-valued (ternary), rather than binary [6].

In this paper, we consider quantum transforms implemented with non-binary quantum circuits as a way to extend the quantum machine utility. While we primarily focus on the implementation with ternary quantum gates, the first motivation is to extend the state space with the given number of quantum digits. Further, due of the apparent inability to realize infinite-precision quantum gates, the approximations of quantum functions are useful [10] and likely inevitable [3]. We show that the proposed transform has better approximation properties than QFT implemented with the binary gates. We further both generalize and improve upon known error bounds [5] thanks to considering more than just a binary case.

The paper first reviews the quantum computing in Section II, followed by the QFT derivations in Section III. In Section IV, we develop ternary quantum transforms and demonstrate their approximation capabilities.

## II. Preliminaries – Quantum States and Gates

Quantum algorithms use the premises of quantum mechanics to achieve parallelism beyond the reach of classical computers. A quantum system realizing such algorithms operates in the Hilbert vector space - we can treat it as a vector space over complex numbers in which the *inner product* over vectors *v* and *w* is defined as:

$$<v|w> = \sum_{i=1}^{n} v_i^* w_i$$

where * denotes complex conjugate. The *norm* of a vector *v* is given as $\|v\| = \sqrt{<v|v>}$. The *orthonormal*

K. Radecka is with Concordia University, Montreal, QC, Canada. (e-mail: kasiar@ece.concordia.ca).

*basis* of the space with *n* quantum basis states $\{|x_1\rangle, |x_2\rangle, \cdots, |x_n\rangle\}$ provides that the inner product of each vector by itself is 1; otherwise it is 0.

The *quantum state* is a vector, i.e., a linear combination of the basic states: $\alpha_1|x_1\rangle + \alpha_2|x_2\rangle + \cdots + \alpha_n|x_n\rangle$. The complex-valued amplitudes $\alpha_i$ are *wave functions* with respect to basis $|x_1\rangle, |x_2\rangle, \cdots, |x_n\rangle$. The *system evolution* in a quantum system is performed through a linear mapping that preserves the vector norm. Hence, each system evolution that transforms a state $\alpha_1|x_1\rangle + \alpha_2|x_2\rangle + \cdots + \alpha_n|x_n\rangle$ into a state

$$\alpha'_1|x_1\rangle + \alpha'_2|x_2\rangle + \cdots + \alpha'_n|x_n\rangle$$

can be expressed via unitary matrix $U$ by $\vec{\alpha}' = U\vec{\alpha}$. As a consequence, such evolutions are always *reversible*. Basic units of information are defined as follows.

**Definition 1:** *A quantum bit, or qubit, is a binary quantum system, defined over the basis $\{|0\rangle, |1\rangle\}$. A q-ary quantum digit is a multiple-valued logic system over basis $\{|0\rangle, |1\rangle, \cdots, |q-1\rangle\}$.*

In literature, the ternary quantum digit is often referred to as a *qutrit*, and we adopt this notation. The state of a single *q*-valued quantum digit is a vector $c_0|0\rangle + c_1|1\rangle + \cdots + c_{q-1}|q-1\rangle$ where the vector norm is 1:

$$|c_0|^2 + |c_1|^2 + \cdots + |c_{q-1}|^2 = 1$$

Since all operations are linear, the relevant matrix notation is summarized in Table 1. For any unitary matrix $A$, $A^{-1}$ equals its Hermitian conjugate (adjoint) $A^\dagger$, i.e., $AA^\dagger = I$.

TABLE 1: STANDARD QUANTUM OPERATION NOTATION

| | |
|---|---|
| $x^*$ | Complex conjugate: $(a+ib)^* = a - ib$ |
| $A^*$ | Complex conjugate of (all entries of) $A$ |
| $A^T$ | Transpose of matrix $A$ |
| $A^\dagger$ | Hermitian conjugate: $\begin{bmatrix} a & c \\ b & d \end{bmatrix}^\dagger = \begin{bmatrix} a^* & b^* \\ c^* & d^* \end{bmatrix}$ |
| $\langle \alpha | \beta \rangle$ | Inner product of vectors $|\alpha\rangle, |\beta\rangle$ |
| $|\alpha\rangle \otimes |\beta\rangle$ or $|\alpha\rangle|\beta\rangle$ | Tensor product of vectors $|\alpha\rangle, |\beta\rangle$ |

### A. Combining the States – Entanglement

Larger quantum systems are obtained by combining individual states. For vectors $x$ and $y$, the compounded state is a linear combination of the new basis states $|x_i y_j\rangle$. Such a state is *decomposable* if it can be represented as:

$$\sum_{i=1}^{n}\sum_{j=1}^{m}\alpha_{ij}|x_i y_j\rangle = \sum_{i=1}^{n}\sum_{j=1}^{m}\alpha_i \beta_j |x_i\rangle|y_j\rangle = \sum_{i=1}^{n}\alpha_i |x_i\rangle \sum_{j=1}^{m}\beta_i |y_i\rangle$$

Otherwise, the state is *entangled*.

*Example 1: a) Consider a system of two qubits, given as*

$$\frac{1}{2}(|00\rangle + |01\rangle + |10\rangle + |11\rangle) = \frac{1}{\sqrt{2}}(|0\rangle + |1\rangle)\frac{1}{\sqrt{2}}(|0\rangle + |1\rangle).$$

*This system is decomposable, as the actions of each qubit are given in the separate brackets on the right-hand side.*

*b) The system $\frac{1}{\sqrt{2}}(|01\rangle + |10\rangle)$ is entangled. It is impossible to express it in the form involving individual qubits.*

In general, the states are compounded by means of the tensor (Kronecker) product of the basic state spaces. Such a combination is referred to as the *quantum register*. The speedups in quantum computation are often due to parallelism offered by the entanglement. Hence, entanglement is a special new resource in quantum computing.

### B. Binary Quantum Gates

As with classical circuits, quantum operations can be performed by networks of gates. Among a large number of possible quantum gates, we now review only the gates used in quantum Fourier transforms.

For reversibility, each gate must have the same number of inputs and the outputs. First, single-input quantum gates are defined by 2x2 unitary matrices over complex numbers.

*Example 2: Consider the Walsh-Hadamard gate:*

$$W_2 = \frac{1}{\sqrt{2}}\begin{bmatrix} 1 & 1 \\ 1 & -1 \end{bmatrix}.$$

*Its application to state |x> is:* $W_2|x\rangle = \frac{1}{\sqrt{2}}(|0\rangle + (-1)^x|1\rangle)$, *which is easily verified by applying it to states |0> and |1>. This gate is equal to its own inverse, since* $(W_2)^2 = I$.

Binary quantum gates with *n* inputs and *n* outputs are given by unitary matrices with $2^n \times 2^n$ entries. Multiple-input, multiple output gates perform unitary operations over a tensor product of single quantum states. There are several common constructions of multiple qubit gates [15]. The *controlled gate* approach can be applied to any single-qubit gate *G*. The first (control) qubit is left unchanged, while the second qubit is affected by *G* only when the first qubit is |1>.

*Example 3: The phase shift gate performs multiplication:*

$$S_\alpha = \begin{bmatrix} 1 & 0 \\ 0 & e^{i\alpha} \end{bmatrix}.$$

*The controlled phase shift gate* $R_\alpha$ *performs the phase shift if the value of the control qubit is 1. The effect of the circuit is given as:* $|00\rangle \to |00\rangle, |01\rangle \to |01\rangle, |10\rangle \to |10\rangle, |11\rangle \to e^{i\alpha}|11\rangle$.

Other multi-qubit constructions include that of *n* independent single-qubit operations. In that case, the function is given by the Kronecker product of the single-qubit matrices.

## III. QUANTUM TRANSFORMS

The Fourier Transform represents elements over an arbitrary Abelian (commutative) group by an expansion over orthogonal sets of basis vectors. We will first describe the traditional development of quantum transforms in the case of multivariate binary inputs, i.e., $(Z_2)^n$ as well as discrete Fourier transforms performed over binary qubits, in which case the group is $Z_{2^n}$.

### A. Quantum Walsh-Hadamard Transform

The first construction is given in terms of *n*-variable quantum functions over binary digits. Then, for vectors *x* and *y*, QFT equals the quantum Walsh-Hadamard Transform,

$$W_{2^n}|x\rangle = \frac{1}{\sqrt{2^n}} \sum_{y \in \{0,1\}^n} (-1)^{xy} |y\rangle.$$

that we denote shorthand as: $|x\rangle \to \frac{1}{\sqrt{2^n}} \sum_{y \in \{0,1\}^n} (-1)^{xy} |y\rangle.$

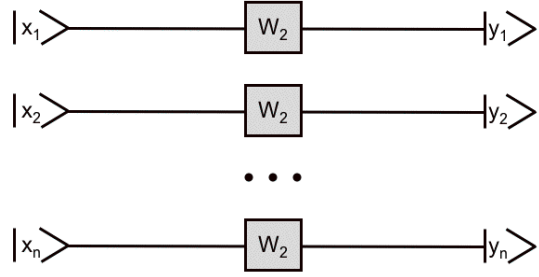

Fig. 1.  Circuit for Quantum Walsh-Hadamard Transform

The transformation of *n*-variable functions is performed by a Kronecker product of univariate transforms. Since the *n*-fold Kronecker product is equivalent to the parallel application of *n* single-qubit functions, the overall transform is performed by only *n* Walsh-Hadamard gates, Fig. 1. The classical fast Walsh Transform requires $O(n2^n)$ operations.

### B. Quantum Discrete Fourier Transform

In general case, Quantum Fourier Transforms defined over a set of *t* values are somewhat more complex to perform. For numbers *x* and *y* the transform has the univariate form:

$$|x\rangle \to \frac{1}{\sqrt{t}} \sum_{y \in \{0,1,\cdots,t-1\}} e^{\frac{-2\pi i x y}{t}} |y\rangle.$$

Unlike previous cases, there is no apparent general way to represent the transform by means of the Kronecker product of the decomposable bases. Similar to the developments in the traditional FFT, several factorizations are possible. We concentrate on the factorization that will allow us to develop ternary transform in the section to follow.

#### 1) Transforming $2^n$ Elements - Qubit Description

The speedup in deriving fast quantum algorithms is due to decomposable realization. The $2^n$ quantities by *n* qubits are represented as $y = y_0 + 2y_1 + 2^2 y_2 + \cdots 2^{n-1} y_{n-1}$ and the decomposable states are: $|y\rangle = |y_0\rangle |y_1\rangle \cdots |y_{n-1}\rangle$. An application of QFT is then:

$$\frac{1}{\sqrt{2^n}} \sum_{y \in \{0,1\}^n} e^{-\frac{2\pi i}{2^n} xy} |y\rangle = \frac{1}{\sqrt{2^n}} \sum_{y \in \{0,1\}^n} e^{-\frac{2\pi i}{2^n} x \sum_{l=1}^{n} y_l 2^l} |y_0 y_1 \cdots y_{n-1}\rangle$$

The key step, analogous to the FFT, is the decomposition $\sum_{y \in \{0,1\}^n} e^{-\frac{2\pi i}{2^n} xy} |y\rangle = \sum_{y' \in \{0,1\}^{n-1}} e^{-\frac{2\pi i}{2^n} x 2 y'} |y' 0\rangle + \sum_{y' \in \{0,1\}^{n-1}} e^{-\frac{2\pi i}{2^n} x(2y'+1)} |y' 1\rangle$ by the least significant bit. Then, QFT is (up to a multiplicative constant) reduced to decomposable states:

$$\sum_{y' \in \{0,1\}^{n-1}} e^{-\frac{2\pi i}{2^n} x 2 y'} |y'\rangle \left( |0\rangle + e^{-\frac{2\pi i x}{2^n}} |1\rangle \right).$$

By using the same construction recursively, QFT reduces to:

$$\left( |0\rangle + e^{\frac{-\pi i x}{2^0}} |1\rangle \right) \left( |0\rangle + e^{\frac{-\pi i x}{2^1}} |1\rangle \right) \cdots \left( |0\rangle + e^{\frac{-\pi i x}{2^{n-1}}} |1\rangle \right). \quad (1)$$

By dividing $x$ with $2^l$, the expression $0.x_{l-1}x_{l-2}...x_0$ denotes the fraction obtained from the least significant $l$ qubits of $x$ as

$$0.x_{l-1}x_{l-2}\cdots x_0 = 2^{-1}x_{l-1} + 2^{-2}x_{l-2} + \cdots + 2^{-l}x_0$$

and the expansion is rewritten into:

$$\left( |0\rangle + e^{-2\pi i 0.x_0} |1\rangle \right) \left( |0\rangle + e^{-2\pi i 0.x_1 x_0} |1\rangle \right) \cdots \left( |0\rangle + e^{-2\pi i 0.x_{n-1}x_{n-2}\cdots x_0} |1\rangle \right).$$

This representation is amenable to the efficient quantum circuit implementations. Each term, multiplied with the appropriate constant, is the unitary transform, built using the Walsh-Hadamard and the controlled rotation gates. The leftmost product term is by definition equal to the function performed by the Walsh-Hadamard gate

$$\left( |0\rangle + e^{-2\pi i 0.x_0} |1\rangle \right) = \left( |0\rangle + e^{-\pi i x_0} |1\rangle \right) = \left( |0\rangle + (-1)^{x_0} |1\rangle \right).$$

The remaining terms are realized following the construction in [5]. For $l^{th}$ term, we first apply the $W_2$ gate, leading to $\frac{1}{\sqrt{2}} \left( |0\rangle + (-1)^{x_{l-1}} |1\rangle \right)$. For each $k < l$, we then perform a phase shift for $\exp(\frac{-\pi i x_k}{2^{l-k}})$, conditional on $k^{th}$ qubit being nonzero. Recalling Example 3, this operation is performed by the controlled phase shift gate $R_k$. Since one phase shift gate is employed for each control qubit $x_k$, $k < l$, this construction is leading up to $n$-1 phase shift gates in total for each qubit. The overall transform can hence be implemented with $n*(n+1)/2$ binary quantum gates, including $n$ Walsh-Hadamard gates.

## IV. QFT USING TERNARY QUANTUM GATES

Since the most pressing limitation in making quantum machines is in the number of quantum digits that can be physically realized, using multi-valued digit representation has its certain advantage in expanding the state space. To realize in $q$-valued case the same number of states as the quantum machine implemented by binary quantum gates, we achieve the reduction by a factor of $\log_2(q)$ qubits. While the proposed implementations are presented here for 3-valued qutrits, the results hold for larger $q$ as well.

### A. Q-ary Quantum Gates

Although the *q*-valued quantum states (*q*>2) come quite natural in many cases [13], there exist few explicit constructions of *q*-valued gates. In developing the representation of *q*-valued states, we find useful the generalization of the Walsh-Hadamard gate, which we call Chrestenson gate after Chrestenson Transform [11].

*Example 4: For the case q=3, the complex 3$^{rd}$ root of unity is:* $a = e^{-\frac{2\pi i}{3}} = \cos(\frac{-2}{3}\pi i) + i\sin(\frac{-2}{3}\pi i) = -0.5 - i*0.866$.
*The Chrestenson gate performs the mapping given by:*

$$CH = \frac{1}{\sqrt{3}}\begin{bmatrix} 1 & 1 & 1 \\ 1 & a & a^2 \\ 1 & a^2 & a \end{bmatrix}.$$

*The adjoint of CH is given as:*

$$\begin{bmatrix} 1 & 1 & 1 \\ 1 & a & a^2 \\ 1 & a^2 & a \end{bmatrix}^\dagger = \begin{bmatrix} 1 & 1 & 1 \\ 1 & a^* & (a^2)^* \\ 1 & (a^2)^* & a^* \end{bmatrix} = \begin{bmatrix} 1 & 1 & 1 \\ 1 & a^* & a \\ 1 & a & a^* \end{bmatrix}.$$

*The gate is unitary, which can be verified by multiplying it by its Hermitian conjugate. By performing such a multiplications and applying identities $a^3=1$ and $1+a+a^2=0$, we obtain the identity matrix.*

Multi-valued quantum gates used in 3-valued implementation include a controlled phase shift gate, which applies to the incoming signal the multiplication by a factor $e^{(i\alpha x)}$, where *x* is the input controlling the amount of shift. Unlike the binary case, there are 3 possible amounts that the signal is phase-shifted by, depending on the value of *x*.

### B. Quantum Chrestenson Transform

The Chrestenson Transform in the *q*-valued case is given by means of orthogonal expansion:

$$|x\rangle \rightarrow \frac{1}{\sqrt{q^n}} \sum_{y \in \{0,1,\cdots,q-1\}^n} e^{\frac{-2\pi i x y}{q}} |y\rangle.$$

The multipliers are equal to $q^{th}$ complex roots of the unity. Please note that by setting *q*=2, we obtain the quantum Walsh-Hadamard Transform as a special case.

For practical calculation, the input and output vectors to the Chrestenson Transform are expressed in terms of their *q*-ary expansions:

$$x = \sum_{i=1}^{n} x_i q^i, \quad y = \sum_{i=1}^{n} y_i q^i,$$

and the decomposable states are: $|x\rangle = |x_0\rangle|x_1\rangle|x_2\rangle \cdots |x_{n-1}\rangle$ and $|y\rangle = |y_0\rangle|y_1\rangle \cdots |y_{n-1}\rangle$.

The transform matrix is then the *n*-fold Kronecker product of univariate transforms:

$$CH_n = CH \otimes CH \otimes \cdots \otimes CH.$$

Further, the quantum Chrestenson Transform can be written in the recursive form. In the case when $q=3$, and by recalling that the third root of unity is $a = e^{-\frac{2\pi i}{3}}$, we obtain the following matrix for the *n*-variable Chrestenson Transform:

$$CH_n = \begin{bmatrix} CH_{n-1} & CH_{n-1} & CH_{n-1} \\ CH_{n-1} & aCH_{n-1} & a^2 CH_{n-1} \\ CH_{n-1} & a^2 CH_{n-1} & aCH_{n-1} \end{bmatrix}.$$

Similarly to the Walsh-Hadamard case, the transform is performed in parallel by *n* Chrestenson gates. We note that the implementation primitives have been elaborated in [13] in terms of ion-trap quantum devices, where it was shown that the implementation using multiple-valued gates fits naturally the technology.

*C. QFT Using Ternary Gates*

We now present the Quantum Fourier Transform over $3^n$ elements implemented using ternary quantum circuits. Analogously to the binary case in Sec. III.B.1) the QFT factorization is conducted as follows [14]:

$$\begin{aligned} QFT_n |x\rangle &= \frac{1}{\sqrt{3^n}} \sum_{y \in \{0,1,2\}^n} e^{\frac{-2\pi i xy}{3^n}} |y\rangle = \\ &= \frac{1}{\sqrt{3^n}} \sum_{y \in \{0,1,2\}^n} e^{-\frac{2\pi i}{3^n} x \sum_{l=1}^n y_l 3^l} |y_0 y_1 \cdots y_{n-1}\rangle = \\ &= \frac{1}{\sqrt{3^n}} \otimes_{l=0}^{n-1} \left[ |0\rangle + e^{-2\pi i x 3^{-l}} |1\rangle + e^{-4\pi i x 3^{-l}} |2\rangle \right] = \\ &= \frac{1}{\sqrt{3^n}} \left( |0\rangle + e^{-2\pi i 0.x_0} |1\rangle + e^{-4\pi i 0.x_0} |2\rangle \right) \cdots \\ &\quad \cdots \left( |0\rangle + e^{-2\pi i 0.x_{n-1} \cdots x_1 x_0} |1\rangle + e^{-4\pi i 0.x_{n-1} \cdots x_1 x_0} |2\rangle \right) \end{aligned} \quad (2)$$

In the final form we employ in this case the ternary fractional number notation:

$$0.x_{l-1} x_{l-2} \cdots x_0 = 3^{-1} x_{l-1} + 3^{-2} x_{l-2} + \cdots + 3^{-l} x_0.$$

The actual circuit for computing the transform is derived by realizing each bracket in Equation (2) with ternary quantum gates. Starting with the leftmost bracket, the whole expression

$$T(x_0) = \frac{1}{\sqrt{3}}\left(|0\rangle + e^{-2\pi i 0.x_0}|1\rangle + e^{-4\pi i 0.x_0}|2\rangle\right)$$

can be realized by a single linear and unitary gate. By considering all three possible values for $|x_0\rangle$ we obtain exactly the action of the Chrestenson gate, and its matrix representation is the same as matrix CH given in Sec. A.

In the final step of deriving the transform, we expand $x$ in terms of its ternary encoding. Considering the term with $3^l$ in the denominator, we express $exp(-2\pi ix/3^l)$ from Equation (2):

$$\exp\left(\frac{-2\pi i\left(3^{n-1}x_{n-1} + 3^{n-1}x_{n-1} + \cdots 3^l x_l + \cdots + 3x_1 + x_0\right)}{3^l}\right) =$$

$$= \exp(-2\pi iK)\exp(\frac{-2\pi i x_{l-1}}{3})\cdots\exp(\frac{-2\pi i x_1}{3^{l-1}})\exp(\frac{-2\pi i x_0}{3^l}).$$

Since $K$ in this expression is an integer, the first term is always equal to 1. Hence, $x_l$ and higher terms in the ternary expansion of $x$ are omitted.

The rest of the QFT circuit is constructed by implementing each term of this expanded exponential function. Please note that for each ternary digit (qutrit) $x_r$, the superposition

$$|0\rangle + e^{-2\pi i x_r 3^{-l}}|1\rangle + e^{-4\pi i x_r 3^{-l}}|2\rangle$$

is realized by the composition of a Chrestenson gate, followed by the phase shift, controlled by $x_k$ ($k<l$):

$$\exp(\frac{-2\pi i x_k}{3^{l-k+1}}).$$

Such controlled shifts involve the least significant digits in the fractional expansion of $x$.

Hence, the overall QFT is performed as in Fig. 2. This implementation consists of one Chrestenson gate per quantum digit, followed by the controlled phase shift gates $R_k$, for a total of $n*(n+1)/2$ gates. Please note that the order of digits at the output is reversed.

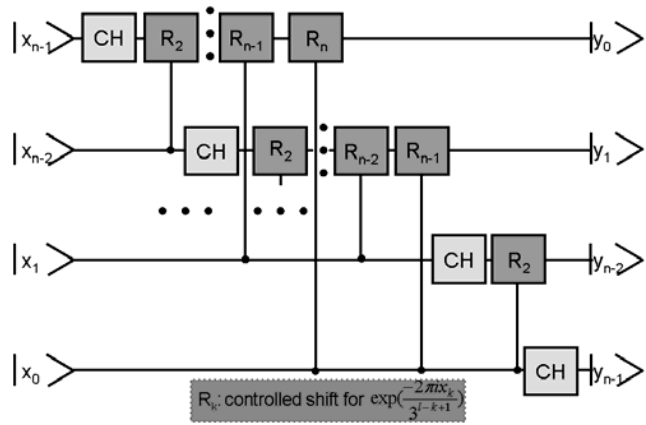

Fig. 2. Circuit for QFT by Q-ary Quantum Gates

*D. Approximation Properties*

The given realization of QFT has useful approximation properties [5]. To obtain a reduced complexity circuit, an approximate construction is derived as follows.

*1) Omitting Controlled Phase Shift Gates*

The controlled phase shift gates in Fig. 2 perform shifts by exponentially decreasing quantities. Therefore, for *q*-valued case, omitting the least significant shift results in a multiplicative factor with error magnitude of

$$e_s = \exp(\frac{-2\pi i(q-1)}{q^l})$$

since in the worst case, the phase shift will be the largest when the control bit is equal to *q*-1.

The magnitude of the multiplicative error can be estimated by its Taylor-Maclaurin expansion:

$\exp(\frac{-2\pi i(q-1)}{q^l}) = 1 - i\frac{2\pi(q-1)}{q^l} - \frac{4\pi^2(q-1)^2}{2!*q^{2l}} + i\frac{8\pi^3(q-1)^3}{3!*q^{3l}} + \cdots$ which is rapidly converging to 1, as the higher-order coefficients decrease exponentially in *q*. Please note that the same error expression holds for binary case, where we obtain:

$$\exp(\frac{-2\pi i}{2^l}) = 1 - i\frac{2\pi}{2^l} - \frac{4\pi^2}{2!*2^{2l}} + i\frac{8\pi^3}{3!*2^{3l}} + \cdots$$

Hence, the convergence towards 1 is the slowest among all values of *q*, and multiple-valued implementations of QFT have better approximation properties.

To estimate the phase of the error due to the omitted controlled phase shift gates, the representation of $e_m$

by trigonometric functions is used to easily separates real and imaginary parts.

$$e_m = \cos(\frac{2\pi(q-1)}{q^l}) - i\sin(\frac{2\pi(q-1)}{q^l})$$

Since the argument in both functions tends towards 0 as $l$ increases, the multiplicative error factor is hence close to the real value of 1, and the phase tends to 0. In comparison with the binary case, where the denominator is equal to $2^l$, the convergence towards 1 is always faster.

*2) New Bounds for Phase Error*

Such good approximation properties allow us to omit more than one controlled phase shift. In general, if $m$ least significant bits are omitted, the overall error exponent is obtained as:

$$\frac{-2\pi i(q^m x_m + \cdots + q\, x_1 + x_0)}{q^l}. \tag{3}$$

The magnitude of the error phase can be determined by the technique in [5] for the approximation error bounds of binary QFT implementation:

$$\frac{2\pi}{q^l} mq^m(q-1) = \frac{2\pi m}{q^{l-m}}(q-1) < \frac{2\pi m}{q^{l-m-1}}.$$

However, as $m$ increases, we now show that even closer bounds are obtained by noticing that the summation of the terms in the bracket of Equation (3) yields a closed-form expression:

$$q^m x_m + \cdots + q\, x_1 + x_0 \le (q-1)\frac{q^m - 1}{q-1} = q^m - 1$$

and the magnitude becomes bounded by $\frac{2\pi}{q^l}(q^m - 1) < \frac{2\pi}{q^{l-m}}$.

This new bound is tighter by a factor of $q*m$. Again, by setting $q=2$ we infer that the binary implementation has the worst approximation properties. For our ternary implementation, the error magnitude is hence reduced by factor of $(3/2)^m$ relative to the binary case.

Note that the given error bound depends on the output digit position, and is the largest for $|y_0\rangle$. The least significant digit has the lowest weight in $q$-ary description of the overall output vector. The most significant digit, however, does not suffer any imprecision due to the omission of the controlled phase shift gates.

## V. Conclusions and Future Work

The Quantum Fourier Transform is the key algorithm in quantum computing. In this paper we presented the implementation of QFT using multiple-valued quantum gates that can overcome two most serious limitations in realizing practical quantum computing machines.

First, since it is currently hard to realize physical systems with large numbers of entangled qubits, the $q$-valued approach reduces the requirements for number of qubits by a factor of $\log_2(q)$. Alternatively, for systems employing $n$ qubits, it expands state space by a factor of $(q/2)^n$.

Second, the proposed implementation possesses better approximation properties than the binary one. By eliminating gates performing the smallest controlled phase shifts, the error magnitude decreases exponentially with $q$. We have also devised new bounds for the error in approximating QFT that is tighter than the original [5] by a factor of $q*m$.

More work can be done in the design of circuits consisting of multiple-valued gates, especially following the results showing that such quantum operations might need to be expressed in terms of Fourier Transform [9]. Binary implementations optimized for the circuit depth have been proposed in [3]. It is worth investigating if such techniques are applicable to ternary quantum circuits. Furthers, the issue of suitable approximation of QFT in such that the decoherence problem gets alleviated, similar to [18] is another worthwhile addition to this study.